\documentclass[
twocolumn,
preprintnumbers,
nofootinbib,
cite,
floats,
prl,
aps
]{revtex4-1}

\usepackage{graphicx}
\usepackage{amsmath, amssymb}
\usepackage{slashed}
\usepackage[T1]{fontenc}

\begin{document}

\preprint{KANAZAWA-18-01}

\title{Superconformal Subcritical Hybrid  Inflation}

\author{Koji Ishiwata}
\affiliation{Institute for Theoretical Physics, Kanazawa University,
  Kanazawa 920-1192, Japan }

\date{\today}

\begin{abstract}
  \noindent
  We consider D-term hybrid inflation in the framework of
  superconformal supergravity. In part of the parameter space,
  inflation continues for subcritical inflaton field value.
  Consequently, a new type of inflation emerges, which gives
  predictions for the scalar spectral index and the tensor-to-scalar
  ratio that are consistent with the Planck 2015 results. The
  potential in the subcritical regime is found to have a similar
  structure to one in the simplest class of superconformal $\alpha$
  attractors.
 \end{abstract}

\maketitle

\section{I. introduction}

The observations of the cosmic microwave background (CMB) strongly
support inflation as the paradigm of early universe. To discover the
nature of inflation, intensive analysis of the CMB has been performed.
The latest results by the Planck
collaboration~\cite{Ade:2015lrj,Ade:2015xua} provide the bounds on the
scalar spectral index $n_s$ and the tensor-to-scalar ratio $r$ of the
primordial density fluctuations,
\begin{align}
  &n_s=0.9655\pm0.0062\,\,
  (68\%\,{\rm CL}) \,,
  \nonumber \\
  &r<0.10\,\, (95\%\,{\rm CL})\,.
  \label{eq:nsr_obs}
\end{align}
  In fact, some inflation models,
such as canonical chaotic inflation~\cite{Linde:1983gd} and hybrid
inflation~\cite{Linde:1993cn}, are already disfavored due to the
bounds.  Although they are not supported by the current observations,
the models are simple and still attractive in theoretical point of
view.

Recently Refs.\,\cite{Buchmuller:2012ex,Buchmuller:2013zfa} studied
the hybrid inflation in the framework of superconformal
supergravity~\cite{Einhorn:2009bh,Kallosh:2010ug,Ferrara:2010yw,Ferrara:2010in}.
It was found that the Starobinsky model~\cite{Starobinsky:1980te}
emerges in the supersymmetric D-term hybrid
inflation~\cite{Binetruy:1996xj,Halyo:1996pp,Kallosh:2003ux}, to give
a good accordance with the Planck observations.  On the other hand,
the D-term hybrid inflation was considered in a different context. In
a shift symmetric K\"{a}hler potential~\cite{Kawasaki:2000yn}, a
`chaotic regime' was found in the subcritical value of the inflaton
field~\cite{Buchmuller:2014rfa}. In the framework, inflation lasts
even after the critical point of the hybrid inflation to give rise to
different predictions from chaotic inflation. The following
study~\cite{Buchmuller:2014dda} showed that the energy scale of
inflation coincides with the Grand Unification (GUT) scale using the
Planck 2013 data~\cite{Planck:2013jfk}.  However, there is a tension
between the predictions and the observations, especially the Planck
2015 data~\cite{Ade:2015lrj,Ade:2015xua}.

In this letter we revisit D-term hybrid inflation in superconformal
framework. It will be shown that there exists a single slow-rolling
field in the subcritical value of the inflaton field. Since inflation
continues for sufficiently long period, cosmic strings are
unobservable as in
Refs.\,\cite{Buchmuller:2014rfa,Buchmuller:2014dda}. The potential in
the subcritical region turns out to be in a general class of
superconformal $\alpha$
attractors~\cite{Kallosh:2013yoa,Kallosh:2013xya}, especially similar
to the simplest version of the model. Consequently, non-trivial
behavior and different predictions from the simplest ones are
discovered.

\section{II. subcritical regime in superconformal D-term inflation}

We consider D-term hybrid inflation in supergravity with
superconformal matter~\cite{Buchmuller:2012ex,Buchmuller:2013zfa}.  In
the model three chiral superfields $S_\pm$ and $\Phi$, which have
local U(1) charge $\pm q$ ($q>0$) and $0$, respectively, are
introduced.  The superpotential and K\"{a}hler potential after fixing
a gauge for the local conformal symmetry are respectively given by,
\begin{align}
  & W=\lambda S_+ S_- \Phi \,, \\
  & K=-3\log \Omega^{-2}\,,
  \label{eq:Kahler}
\end{align}
with,
\begin{eqnarray}
  \Omega^{-2}=1-\frac{1}{3}\left(|S_+|^2+|S_-|^2+|\Phi|^2\right) -
  \frac{\chi}{6}\left(\Phi^2+\bar{\Phi}^2\right)\,,
  \nonumber \\
\end{eqnarray}
where $\lambda$ and $\chi$ are constants.\footnote{Throughout this
  letter we use the same notation for chiral superfields and scalar
  fields and take the reduced Planck mass $M_{\rm pl}=1$ unit. } The
term proportional to $\chi$ in the K\"{a}hler potential breaks
superconformal symmetry explicitly.  In the model the Fayet-Iliopoulos (FI)
term can be accommodated. Then, the D-term potential in the Einstein
frame is~\cite{Buchmuller:2012ex},
\begin{eqnarray}
V_D=\frac{1}{2}g^2\left(q\Omega^2(|S_+|^2-|S_-|^2)-\xi\right)^2\,,
\end{eqnarray}
where $g$ is the gauge coupling and $\xi$ is the FI term, which is
taken as a constant.  (See
Refs.\,\cite{Binetruy:2004hh,Komargodski:2010rb,Dienes:2009td,
  Catino:2011mu,Wieck:2014xxa,Domcke:2014zqa} for the subtleties of
this issue in supergravity.) The F-term potential in the Einstein
frame, on the other hand, is given in a simple form without
exponentially growing terms~\cite{Buchmuller:2012ex,Einhorn:2012ih},
\begin{align}
  V_F &=\Omega^{4} \lambda^2\biggl[
    |\Phi|^2\left(|S_+|^2+|S_-|^2\right)+|S_+S_-|^2 \nonumber \\
&    
    -\frac{\chi^2|S_+S_-\Phi|^2}
    {3+\frac{\chi}{2}\left(\Phi^2+\bar{\Phi}^2\right)+\chi^2|\Phi|^2} \biggr]\,.
\end{align}
As in the canonical hybrid inflation, $S_-$ is stabilized to its
origin meanwhile $S_+$ suffers from the tachyonic instability
depending on the field value of $\Phi$. The nature of $\Phi$ depends
on the value of $\chi$. In the K\"{a}hler potential there is a shift
symmetry under ${\rm Re}\,\Phi \,({\rm Im}\,\Phi) \to {\rm Re}\,\Phi\,
({\rm Im}\,\Phi)\,+$\,const. for $\chi=-1\,(+1)$, and ${\rm Re}\,\Phi
\,({\rm Im}\,\Phi)$ can play a role of inflaton, as mentioned in
Ref.\,\cite{Buchmuller:2012ex}. We consider $\chi \le -1$ in the later
discussion without loss of generality. Then, the total potential is
given by the waterfall field $s\equiv \sqrt{2}|S_+|$ and the inflaton
field $\phi\equiv \sqrt{2}{\rm Re}\,\Phi$,
\begin{align}
  V_{\rm tot}(\phi,s)&=V_F+V_D \nonumber \\
  &=\frac{\Omega^4(\phi,s) \lambda^2}{4} s^2 \phi^2
  + \frac{g^2}{8}\left(q \Omega^2(\phi,s)s^2-2\xi\right)^2\,, \\
  \Omega^{-2}(\phi,s)&=1-\frac{1}{6}\left(s^2+(1+\chi)\phi^2\right)\,.
\end{align}
The waterfall field becomes tachyonic below the critical value
$\phi_c$ of the inflaton field,
\begin{align}
  \phi_c^2=\frac{6qg^2\xi}{3\lambda^2+(1+\chi)qg^2\xi}\,.
\end{align}
After the tachyonic growth, the waterfall field is expected to reach
its local minimum, which is obtained by $\partial V_{\rm
  tot}(\phi,s)/\partial s =0$,
\begin{align}
  s_{\rm min}^2
  &=\frac{2\xi\Omega^{-2}(\phi,0)}{q(1+\tilde{\xi})}
  \frac{1-\Psi^2}{1+\frac{\tilde{\xi}}{1+\tilde{\xi}}\Psi^2}\,,
  \label{eq:smin}
\end{align}
where $\tilde{\xi}\equiv \xi/3q$ and,
\begin{align}
  \Psi\equiv \frac{\Omega(\phi,0)\phi}{\Omega(\phi_c,0)\phi_c}
=\frac{\Omega(\phi,0)\phi}{\sqrt{2qg^2\xi/\lambda^2}}\,.
\end{align}
The expression for the local minimum given in
Refs.\,\cite{Buchmuller:2014rfa,Buchmuller:2014dda} corresponds to the
case for $\chi=-1$ (and $q=1$) from the facts that
$\Omega(\phi,0)|_{\chi=-1}=1$ and $\tilde{\xi}\sim {\cal O}(10^{-4})$
in our targeted parameter space. Following
Refs.\,\cite{Buchmuller:2014rfa,Asaka:2001ez} (see also Appendix), we
have confirmed numerically that the waterfall field reaches to the
local minimum after ${\cal O}(1/H_c)$ where $H_c(=g\xi/\sqrt{6})$ is
the Hubble parameter at the critical point, and then it becomes a
single field inflation.  Since the inflation lasts well over ${\cal
  O}(10^{2}/H_c)$, cosmic strings, which are produced during the
tachyonic growth, are unobservable.  After the waterfall field relaxed
to the local minimum, the dynamics of the inflaton is described by the
potential,
\begin{align}
  V&\equiv V_{\rm tot}(\phi,s_{\rm min}) \nonumber \\
  &=  g^2\xi^2(1+\tilde{\xi})\Psi^2
        \frac{1-\frac{\Psi^2}{2(1+\tilde{\xi})}}
        {1+2\tilde{\xi}\Psi^2} \,.
        \label{eq:V}
\end{align}
As in Eq.\,\eqref{eq:smin}, it is easily to see that the potential $V$
with $\chi=-1$ agrees with one given in
Refs.\,\cite{Buchmuller:2014rfa,Buchmuller:2014dda} up to ${\cal
  O}(\xi)$.\footnote{$q$ and $g$ can be absorbed by the redefinition
  of $\lambda$ and $\xi$, $\bar{\lambda}\equiv \lambda /\sqrt{qg}$ and
  $\bar{\xi}\equiv g \xi$ if we ignore terms proportional to
  $\tilde{\xi}$ that are irrelevant numerically.  Although we will use
  $\lambda$ and $\xi$ in the following discussion, the results in
  terms of $\bar{\lambda}$ and $\bar{\xi}$ can be obtained by $q\to 1$,
  $g\to 1$, $\lambda \to \bar{\lambda}$, and $\xi \to \bar{\xi}$.}

We note that non-zero $\lambda$ explicitly breaks the shift symmetry
for ${\rm Re}\,\Phi$ as well as $\chi$ that deviates from $-1$
does. Thus, a parameter $\lambda \ll 1$ and $\chi\simeq-1$ is
consistent with each other under the approximate shift symmetry. In
addition, $\chi\simeq -1$ is required for $\lambda \ll 1$ otherwise
$\phi^2_c$ gets negative.  As it will be seen, the observational data
indeed implies such a parameter space.

\section{III. cosmological consequences}

\begin{figure}[t]
  \begin{center}
    \includegraphics[scale=0.5]{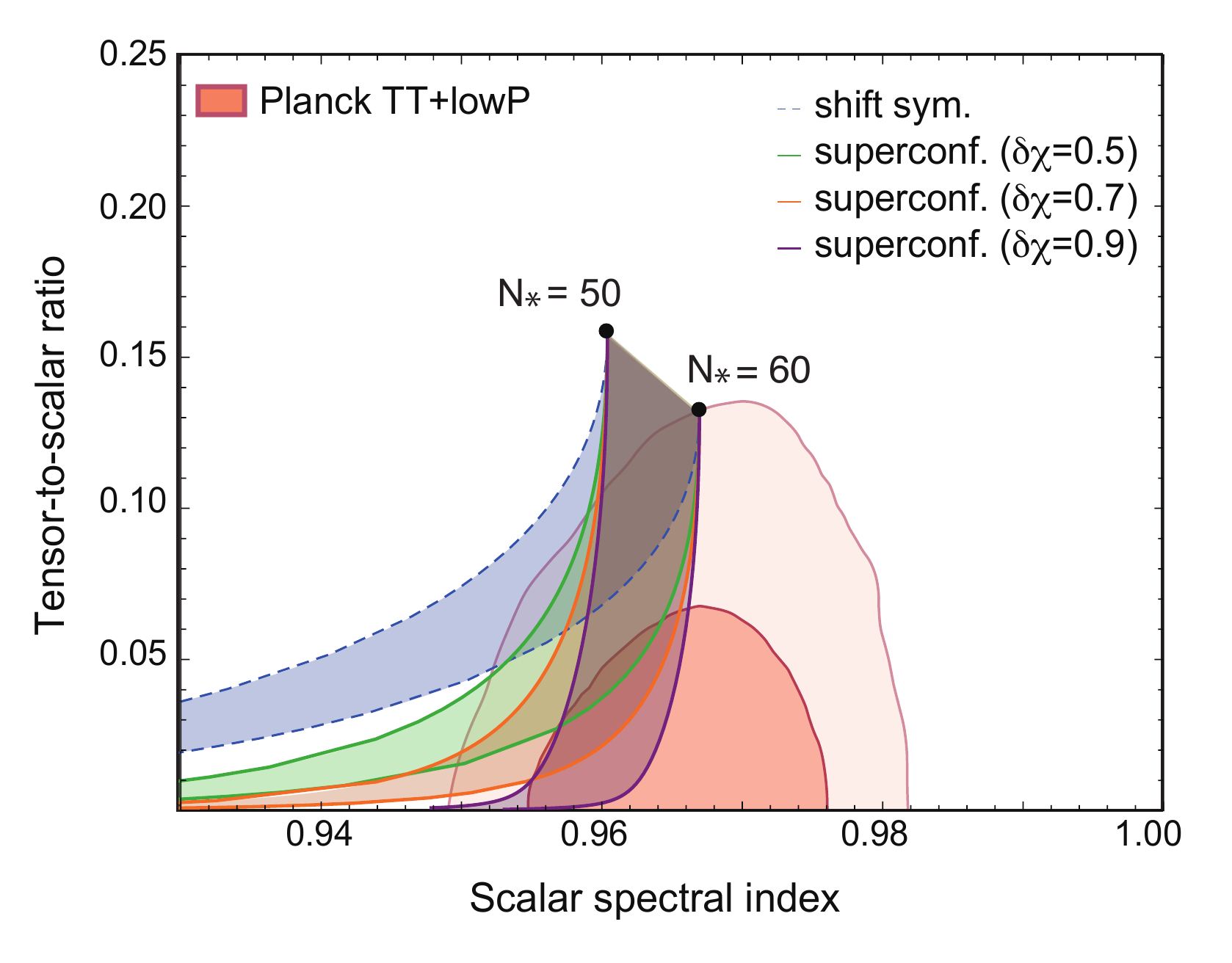}
  \end{center}
  \caption{Scalar spectral index and tensor-to-scalar ratio for
    various values of $\delta\chi$ ($0<\delta\chi<1$) and $N_{*}=50$
    and $60$ as `superconf.'. Result in the shift symmetric K\"{a}hler
    potential in Ref.\,\cite{Buchmuller:2014dda} is also given as
    `shift sym.' (updated using the Planck 2015 results). Here $q=g=1$
    and appropriate values of $\lambda$ and $\xi$ to satisfy the
    observed scalar amplitude is taken. }
  \label{fig:nsr}
\end{figure}

The slow roll parameters for the inflaton dynamics are given as,
\begin{align}
  \epsilon(\phi) = \frac{1}{2}\left(\frac{V'}{V}\right)^2\,,
  \quad \quad
  \eta(\phi) =\frac{V''}{V}\,,
\end{align}
where $V'=dV/d\hat{\phi}$ and $V''=d^2V/d\hat{\phi}^2$. Here
$\hat{\phi}$ is canonically-normalized inflaton field that is related
to $\phi$ as,
\begin{align}
  \frac{d \phi}{d\hat{\phi}}=K_{\Phi \bar{\Phi}}^{-1/2}\,,
  \label{eq:dphidphihat}
\end{align}
where $K_{\alpha\bar{\alpha}}\equiv\partial^2
K/\partial\alpha\partial{\bar{\alpha}}$. $|S_-|=0$,
$\Phi=\bar{\Phi}=\phi/\sqrt{2}$, and $|S_+|=s_{\rm min}/\sqrt{2}$ are
implicit here. (Parametrically $s_{\rm min}\simeq 0$ is a good
approximation as discussed later.) Inflation ends at $\phi=\phi_f\equiv{\rm
  Max}\{\phi_{\epsilon}$,\,$\phi_{\eta}$\} where
$\epsilon(\phi_\epsilon)=1$ and $|\eta(\phi_\eta)|=1$, and the last
$e$-folds $N_*$ before the end of inflation is obtained by,
\begin{align}
  N_*=\int_{\phi_f}^{\phi_*}d\phi \frac{V}{dV/d\phi}K_{\Phi\bar{\Phi}}\,.
\end{align}
The cosmological observables, {\it i.e.,} the scalar amplitude $A_s$,
the spectral index, and the tensor-to-scalar ratio, are then
determined by,
\begin{align}
  A_s=&\frac{V(\phi_*)}{24\pi^2 \epsilon(\phi_*)}\,,
  \\
  n_s=1+2\eta(\phi_*)-6\epsilon&(\phi_*)\,, \quad \quad 
  r=16 \epsilon(\phi_*)\,.
\end{align}
We normalize the scalar amplitude by using the Planck 2015
data\,\cite{Ade:2015xua} $A_s=2.198^{+0.076}_{-0.085}\times 10^{-9}$
and compute $n_s$ and $r$ for a given $N_*$.

As we have stated before, our target is the parameter space $\lambda
\ll 1$.  To search such a region, it is convenient to parametrize
$\chi$ as,
\begin{align}
  \chi=-1-\frac{3\lambda^2}{qg^2\xi}\delta \chi \quad \quad
  (0<\delta\chi<1)\,,
\end{align}
to satisfy $\phi_c^2=2qg^2\xi/\lambda^2(1-\delta\chi)>0$. 

\begin{figure}[t]
  \begin{center}
    \includegraphics[scale=0.65]{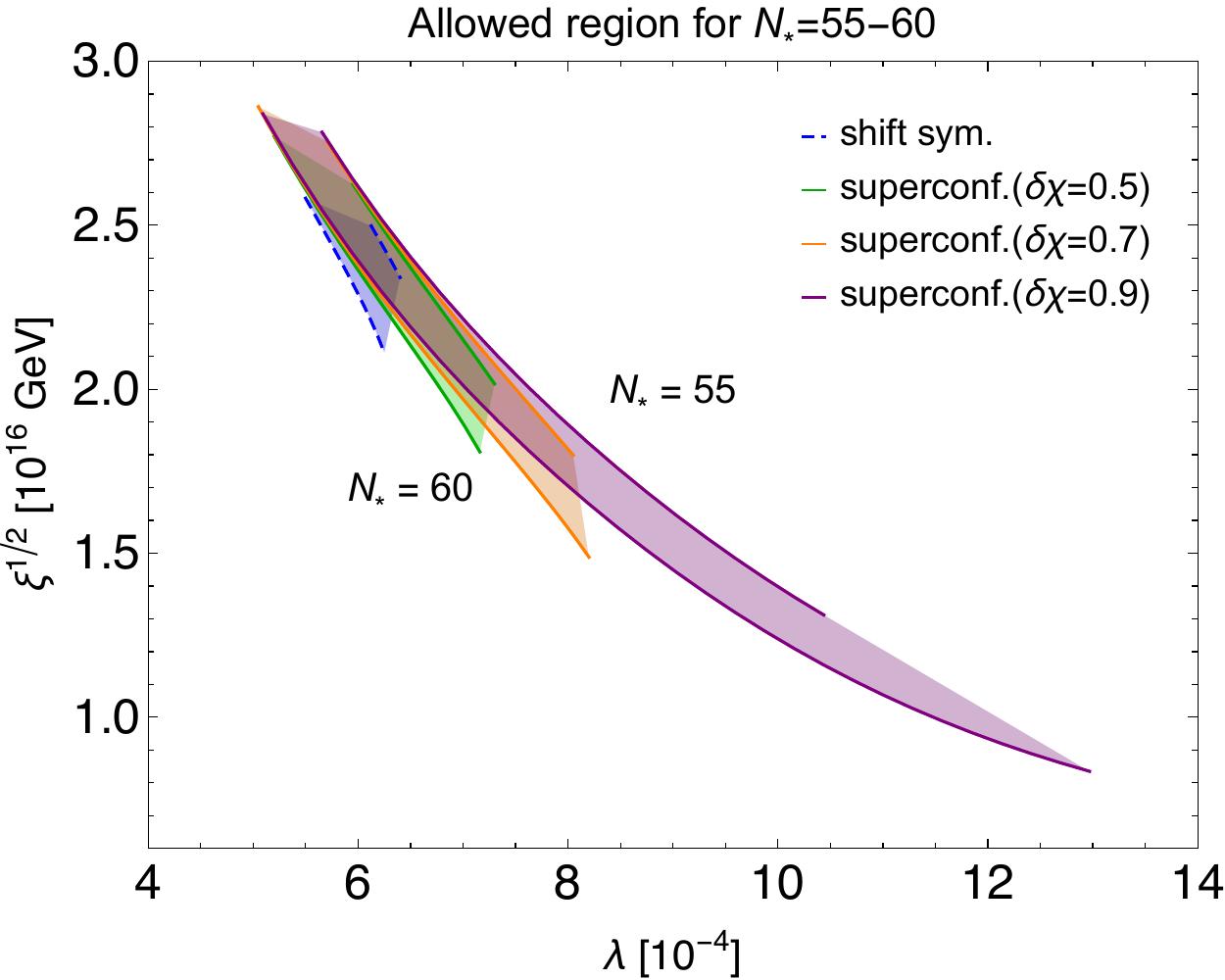}
  \end{center}
  \caption{Allowed region for $N_*=55$ to 60 from the bounds on $n_s$
    (68\% CL) and $r$ (95\% CL). Line contents are the same as
    Fig.\,\ref{fig:nsr}. Here we have updated the result for the shift
    symmetric K\"{a}hler case by using the Planck 2015 data.}
  \label{fig:lamxi}
\end{figure}

Now we are ready to discuss the cosmological
consequences. Fig.\,\ref{fig:nsr} shows the predictions of $n_s$ and
$r$ in our model. Here $q=g=1$ is taken (see footnote 2), and
$\lambda$ and $\xi$ are determined for a $\delta \chi$ and $N_*$ by
using the scalar amplitude observed by the Planck collaboration.  In
Fig.\,\ref{fig:lamxi}, the allowed regions due to the bounds on $n_s$
and $r$ are shown for $N_*=55$--60.\footnote{There is no allowed
  region for $N_*=50$ except for $\delta \chi=0.9$. } The upper and
lower bounds on $\xi$ corresponds to the upper limit on $r$ and lower
limit on $n_s$, respectively. In the $n_s$-$r$ plane, smaller values
of $n_s$ and $r$ are obtained for larger $\lambda$ (and smaller
$\xi$). In Fig.\,\ref{fig:nsr} the result in the previous work
\cite{Buchmuller:2014dda}, {\it i.e.}, the shift symmetric K\"{a}hler
potential case, is also given as `shift sym.'.\footnote{ Do not
  confuse with the shift symmetric K\"{a}hler case with the present
  superconformal case where the shift symmetry is (weakly) broken in
  the K\"{a}hler potential.} We have checked that the result for
$\delta \chi=0$ agrees with it numerically and the similar behavior is
seen around $\delta \chi\simeq 0$. When $\delta \chi$ gets close to
unity, on the contrary, a different behavior is observed.  It is seen
that $r$ gets smaller meanwhile $n_s$ tends to stay in the same value,
which is within the Planck bounds.  As a result, a wider allowed
parameter space is obtained, which is seen in Fig.\,\ref{fig:lamxi}.

It is seen $\lambda\sim 10^{-4}$--$10^{-3}$ and $\sqrt{\xi}\sim
10^{16}$\,GeV are consistent region with the Planck observation.
Although the allowed region becomes larger, $\sqrt{\xi}$ tends to sit
around the GUT scale even for $\delta \chi=0.9$. As a consequence, the
predicted $r$ is not extremely small.  For example, $r>0.0020$ (0.075)
for $N_*=60$ (50) for $\delta \chi=0.9$. The value of $\chi$ in the
allowed region, on the other hand, is found as $-1.41$
$(-1.016)<\chi<-1.0046$ ($-1.0092$) for $N_*=60$ (50). Therefore, the
parameter space $\lambda \ll 1$ and $\chi\sim -1$ is indeed favored by
the observations.

In order to interpret the results, it is instructive to consider a
canonically-normalized inflaton field $\hat{\phi}$. Although the r.h.s
of Eq.\,\eqref{eq:dphidphihat} is complicated, it can be approximated
in the parameter space we are considering as,
\begin{align}
  \frac{d \phi}{d\hat{\phi}}
  \simeq \sqrt{1-\frac{1}{6}(1+\chi)\phi^2}\,.
  \label{eq:dphidphihat_app}
\end{align}
Then it is solved analytically,
\begin{align}
  \phi=\frac{1}{\sqrt{\beta}}\sinh\sqrt{\beta}(\hat{\phi}+C)\,,
  \label{eq:phiinphihat}
\end{align}
where $C$ is a constant and,
\begin{align}
  \beta\equiv -\frac{1+\chi}{6}=\frac{\lambda^2}{2qg^2\xi}\delta\chi
  =\frac{\delta \chi}{\phi^2_c(1-\delta\chi)}\,.
  \label{eq:beta}
\end{align}
We have found that $C=0$ is appropriate choice.  Then $\Psi$ is simply
given as,
\begin{align}
\Psi\simeq \delta \chi^{-1/2} \tanh\sqrt{\beta} \hat{\phi}\,,  
\end{align}
to express the potential in terms of $\hat{\phi}$,
\begin{align}
  V\simeq g^2\xi^2 \delta \chi^{-1} \tanh^2\sqrt{\beta} \hat{\phi}
  \biggr[1-\frac{\delta \chi^{-1}}{2} \tanh^2\sqrt{\beta}
    \hat{\phi}\biggl]\,.
  \label{eq:Vcano}
\end{align}
This potential is valid in $\hat{\phi}\le \hat{\phi}_c=
\frac{1}{\sqrt{\beta}}\sinh^{-1}\sqrt{\beta}\phi_c$. It is
straightforward to check that the r.h.s is equal to $g^2 \xi^2/2$ for
$\hat{\phi}=\hat{\phi}_c$, and $\hat{\phi}_c\to \infty$ for $\delta
\chi\to 1$.  We note that the potential coincides with a general class
of superconformal $\alpha$ attractors~\cite{Kallosh:2013yoa}.  It
especially resembles to the simplest class of the model,
\begin{align}
  V_{\alpha \mathchar`-attr}=
  \Lambda^4 \tanh^{2m}\frac{\hat{\phi}}{\sqrt{6\alpha}}\,.
  \label{eq:Valpha}
\end{align}
Due to the additional term, however, it has a different asymptotic
behavior as we will see below.

In the small $\lambda$ (and large $\xi$) region, $\beta$ gets small,
then the potential reduces to,
\begin{align}
  V\simeq
g^2\xi^2(1-\delta\chi) \frac{\hat{\phi}^2}{\phi_c^2}
\left[1-\frac{1+(4/3)\delta\chi}{2(1-\delta\chi)}
  \frac{\hat{\phi}^2}{\phi^2_c}\right]\,.
\end{align}
This is nothing but the potential for the shift symmetric K\"{a}hler
case given in Refs.\,\cite{Buchmuller:2014rfa,Buchmuller:2014dda} in
the limit $\delta \chi \to 0$, which leads to $\hat{\phi}\to \phi$.
This feature is clearly seen in Fig.\,\ref{fig:nsr}. We note that the
quadratic term is rewritten as $(\lambda^2\xi/2q)\hat{\phi}^2$, which
is independent of $\delta\chi$. Therefore, $n_s$ and $r$ approach to
those in quadratic chaotic inflation in small $\lambda$ limit (while
$\lambda^2 \xi\simeq$ constant), independent of $\delta \chi$.
(Such a region is excluded, thus it is not shown in
Fig.\,\ref{fig:lamxi}.)  The potential $V_{\alpha \mathchar`-attr}$,
on the other hand, has a similar structure,
\begin{align}
  V_{\alpha \mathchar`-attr}\simeq \frac{\Lambda^4}{6\alpha}\hat{\phi}^{2m}
  \left[1-\frac{m\hat{\phi}^2}{9\alpha}\right]\,.
\end{align}
Although it coincides with $V$ in the limit $\hat{\phi}\to 0$ for
$m=1$ and $\Lambda^4/6\alpha=\lambda^2\xi/2q$, it is not possible
to get the same factor for the quartic term.

\begin{figure}[t]
  \begin{center}
    \includegraphics[scale=0.65]{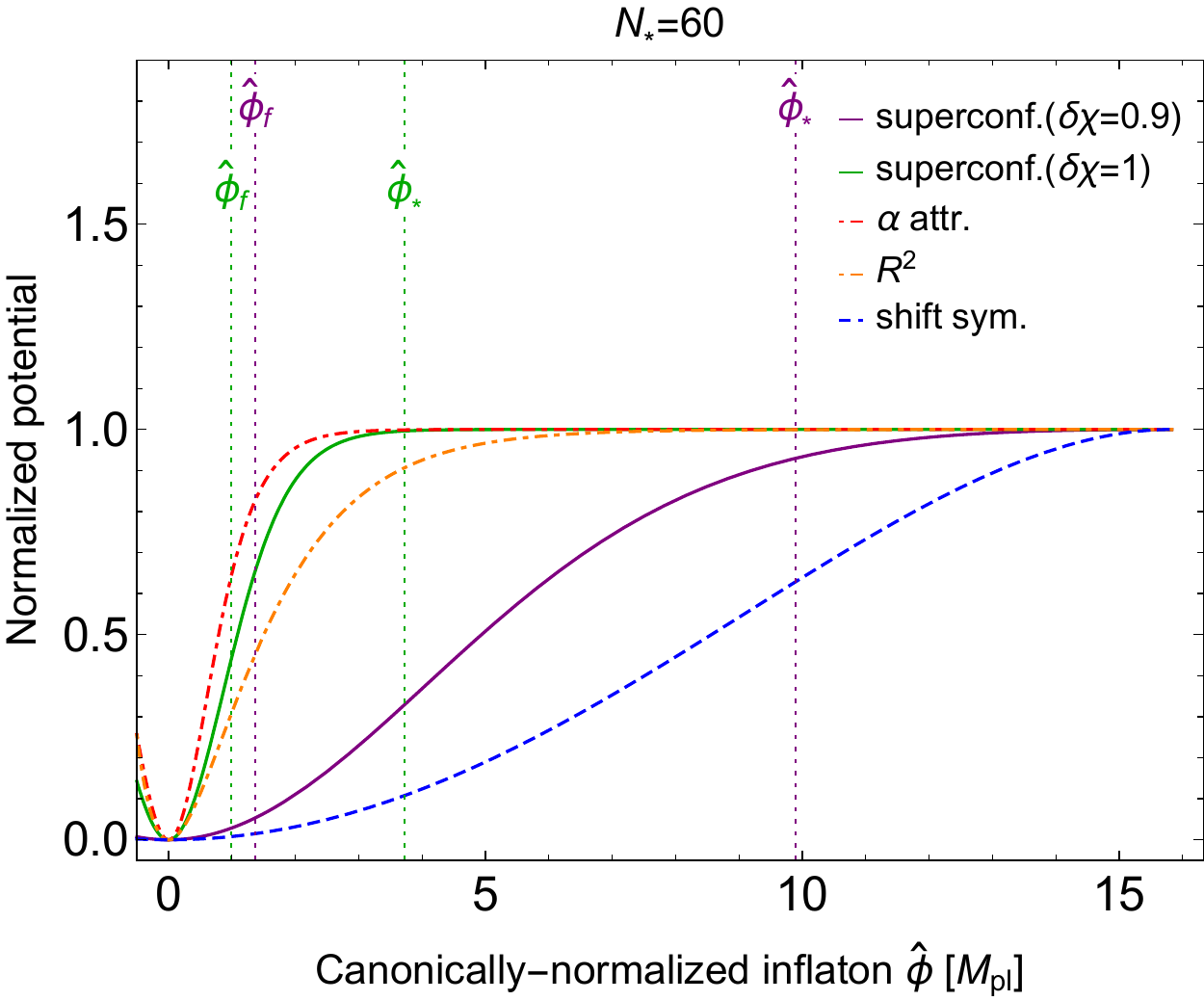}
  \end{center}
  \caption{Potential as function of canonically-normalized inflaton
    field $\hat{\phi}$. $\delta\chi=0.9$ and $1$ cases (`superconf.')
    are shown, which are compared with superconformal $\alpha$
    attractors (`$\alpha$ attr.'),  $R^2$ inflation (`$R^2$'), and
    the shift symmetric K\"{a}hler case (`shift sym.').  The field
    values $\hat{\phi}_f$ and $\hat{\phi}_*$ at the end of inflation
    and the last 60 $e$-folds, respectively, are also indicated for
    $\delta\chi=0.9$ and $1$ cases. }
  \label{fig:V}
\end{figure}

In large $\lambda$ (and small $\xi$) region, on the contrary, $\beta$
increases, which leads us to expand $\Psi$ in large
$\sqrt{\beta}\hat{\phi}$ limit to obtain,
\begin{align}
  V\simeq \frac{1}{2}g^2\xi^2(2-\delta\chi^{-1})
  \left[1+a_1e^{-2\sqrt{\beta}\hat{\phi}}
    -a_2e^{-4\sqrt{\beta}\hat{\phi}}\right]\,,
  \label{eq:Vlargefield}
\end{align}
with $a_1=8(1-\delta\chi)/(2\delta\chi-1)$ and
$a_2=16(2-\delta\chi)/(2\delta\chi-1)$. This expression should be
compared with Eq.\,\eqref{eq:Valpha} in the $\alpha\ll 1$ limit. As
shown in Ref.\,\cite{Kallosh:2013yoa}, it reduces to the potential in
$R^2$ inflation~\cite{Whitt:1984pd} at large field value\footnote{To
  be precise, $\alpha=1$ gives the original $R^2$ inflation. The
  factor $4m (>0)$ is quantitatively irrelevant for the slow-roll
  predictions.},
\begin{align}
  V_{\alpha \mathchar`-attr}\simeq \Lambda^4\left[
    1-4me^{-\frac{2\hat{\phi}}{\sqrt{6\alpha}}}\right]\,.
\end{align}
Now it is clear that the form of the potential with $\delta\chi=1$ (in
large $\lambda$ region) reduces to $R^2$ inflation, or the simplest
class of superconformal $\alpha$ attractors in $\alpha\ll 1$ limit. To
be specific, a choice of $\Lambda^4=g^2\xi^2/2$ and $\alpha=1/24\beta$
leads to the same asymptotic form. Then, we get $n_s\simeq
1-2/(N_*+1)-3qg^2\xi/8\lambda^2(N_*+1)^2$, $r\simeq
qg^2\xi/\lambda^2(N_*+1)^2$, while satisfying $\lambda^2 \xi\simeq$
constant. Namely, when $\lambda$ increases $n_s$ approaches to
$1-2/(N_*+1)$ and $r$ gets smaller and smaller.  We have confirmed
this behavior using Eq.\,\eqref{eq:Vcano} with $\delta\chi=1$.  Recall
that, however, the critical value becomes infinity, which is
unphysical.

Such a behavior, on the contrary, can not be seen for $\delta \chi\neq
1$ case shown in Fig.\,\ref{fig:nsr}. This arises from non-zero $a_1$
in Eq.\,\eqref{eq:Vlargefield}. This is why we have seen the different
cosmological consequences.

In Fig.\,\ref{fig:V}, the potential as function of
canonically-normalized inflaton field is plotted for $\delta \chi=0.9$
and $1$.  Here $\lambda=9.4 \times 10^{-4}$ ($1.9\times 10^{-3}$),
$\sqrt{\xi}=1.3\times 10^{16}$ ($5.7 \times 10^{15}$)\,GeV for $\delta
\chi=0.9$ (1) to give $n_s=0.966$ and $r=0.051$ (0.00052) for
$N_*=60$. In the plot the potentials in superconformal $\alpha$
attractors $V_{\alpha\mathchar`-attr}$, $R^2$ inflation, and the shift
symmetric K\"{a}hler case, are also shown for comparison.  Each
potential is normalized to unity when $\hat{\phi}$ reaches to the
critical point ($\delta \chi=0.9$ case and the shift symmetric
K\"{a}hler case) or infinity ($\delta \chi=1$ case, $\alpha$
attractors, and $R^2$ inflation). For the shift symmetric K\"{a}hler
case, the critical point is taken to the same value as
$\delta\chi=0.9$ case. We take the parameters for $\alpha$ attractors
and $R^2$ inflation to have the same asymptotic form as $\delta
\chi=1$ case in large field limit. It is seen that $\delta \chi=1$
case is similar to $\alpha$ attractors, but not exactly the same.  As
a result, the predictions for the slow-roll quantities are different,
{\it i.e.,} $n_s=0.967$ and $r=0.00044$.  It is clear, on the other
hand, that $\delta \chi=0.9$ case shows a different behavior from the
others. To summarize, the model has a nature of both the shift
symmetric K\"{a}hler case and the simplest superconformal $\alpha$
attractors, and the slow-roll predictions change accordingly.

\section{IV. conclusion}

We have revisited superconformal D-term hybrid inflation. After
reaching its critical value, the inflaton field is slowly rolling thus
inflation continues for a small coupling $\lambda$ of inflaton to the
other fields. Because of a sufficiently long period of slow-roll
regime, cosmic strings, which are formed during the tachyonic growth
of the waterfall field, are unobservable.  The potential which
determines the dynamics of the canonically-normalized inflaton in the
subcritical regime has been found to resemble to the simplest version
of superconformal $\alpha$ attractors but with an additional term.
Consequently, different predictions for the slow-roll parameters are
obtained.  For $\lambda\sim 10^{-4}$--$10^{-3}$ and $\sqrt{\xi}\sim
10^{16}\,{\rm GeV}$, $n_s$ and $r$ are consistent with the Planck
data.

The predictions depend on a parameter $\chi$ that explicitly breaks
superconformal symmetry in the K\"{a}hler potential. In addition, the
K\"{a}hler potential with $|\chi|=1$ has a shift symmetry for the
inflaton field, which is explicitly broken by non-zero $\lambda$ in
the superpotential. On the other hand, $|\chi|\simeq 1$ is required
from the consistency of the model setting, thus $\lambda\ll 1$ is
parametrically natural. It has been found that the observational
bounds indeed prefer such a parameter space.

\vspace{0.2cm}

\noindent
{\it Acknowledgments}\\
\noindent
We are grateful to Wilfried Buchm\"{u}ller for valuable discussions
and helpful comments on the manuscript.  This work was supported by
JSPS KAKENHI Grant Numbers JP17H05402, JP17K14278 and JP17H02875.

\appendix*

\section{Appendix: Tachyonic growth of waterfall field}

Around the critical value of the inflaton field, the dynamics of the
waterfall field is governed by its tachyonic growth. For the
evaluation, we define canonically-normalized waterfall field $\hat{s}$
around the critical point,
\begin{align}
  \frac{ds}{d\hat{s}}\simeq K_{S_+\bar{S}_+}^{-1/2}(\phi_c,0)\,.
\end{align}
The rest parts are parallel to 
Refs.\,\cite{Buchmuller:2014rfa,Buchmuller:2014dda}. We expand the
potential $V_{\rm tot}(\phi,s)$ near the critical point, {\it i.e.,} $\phi\simeq
\phi_c+\dot{\phi}_c t$ as
\begin{align}
  V_{\rm tot}(\phi,s)=
  \frac{g^2\xi^2}{2}-\frac{d^3}{2}t s^2 +{\cal O}(t^2, s^4)\,,
\end{align}
to leads the interaction term in the equation of motion of $\hat{s}$,
\begin{align}
  \frac{\partial V_{\rm tot}(\phi,s)}{\partial \hat{s}}&=
  K_{S_+\bar{S}_+}^{-1/2}(\phi,s) \frac{\partial V(\phi,s)}{\partial s}
  \nonumber \\
  &=-\hat{d}^3\hat{s} +{\cal O}(t^2, s^3)\,,
\end{align}
where $\hat{d}^3=(2qg^2\xi)^2 |\dot{\phi}_c|/\lambda^2\phi_c^3$
and $|\dot{\phi}_c|=-(\partial V_{1l}/\partial \phi)/(3H_c)=
\sqrt{6}\log 2\,
q^2g^3\lambda^2\xi/4\pi^2\phi_c(3\lambda^2+qg^2\xi(1+\chi)^2)$.  The
equation of motion for $\hat{s}$ gives rise to that of momentum mode
$\hat{s}_k$ of the quantum
fluctuation\,\cite{Buchmuller:2014rfa,Buchmuller:2014dda,Asaka:2001ez},
\begin{align}
  \ddot{\hat{s}}_k +
  \Bigl[k^2e^{-2H_ct}-\frac{9}{4}H_c^2-\hat{d}^3t\Bigr]\hat{s}_k=0\ .
\end{align}
Here we have used $V_{1l}$ given in
Ref.\,\cite{Buchmuller:2012ex}. Solving the equation numerically, we
obtain the variance $\langle \hat{s}^2 (t)\rangle$.

After the decoherence time $t_{\rm dec}$ we match the variance with
the classical motion of the waterfall field as $s(t_{\rm
  dec})=K_{S_+\bar{S}_+}^{-1/2}(\phi_c,0)\sqrt{\langle \hat{s}^2
  (t_{\rm dec})\rangle}$, and solve the equations of motion for $\phi$
and $s$
\begin{align}
  &3H\dot{\phi}=-K_{\Phi\bar{\Phi}}^{-1}(\phi,s)
  \frac{\partial V(\phi,s)}{\partial \phi}\,,
  \\
  &3H\dot{s}=-K_{S_+\bar{S}_+}^{-1}(\phi,s)
  \frac{\partial V(\phi,s)}{\partial s}\,.
\end{align}
We have confirmed numerically that the obtained solutions coincide
with the dynamics described by $V$ in Eq.\,\eqref{eq:V}.



\begin{thebibliography}{99}

 \bibitem{Ade:2015lrj} 
  P.~A.~R.~Ade {\it et al.} [Planck Collaboration],
  Astron.\ Astrophys.\  {\bf 594}, A20 (2016)
  doi:10.1051/0004-6361/201525898
  [arXiv:1502.02114 [astro-ph.CO]].

\bibitem{Ade:2015xua} 
  P.~A.~R.~Ade {\it et al.} [Planck Collaboration],
  Astron.\ Astrophys.\  {\bf 594}, A13 (2016)
  doi:10.1051/0004-6361/201525830
  [arXiv:1502.01589 [astro-ph.CO]].

\bibitem{Linde:1983gd} 
  A.~D.~Linde,
  Phys.\ Lett.\  {\bf 129B}, 177 (1983).
  doi:10.1016/0370-2693(83)90837-7

  
\bibitem{Linde:1993cn} 
  A.~D.~Linde,
  Phys.\ Rev.\ D {\bf 49}, 748 (1994)
  doi:10.1103/PhysRevD.49.748
  [astro-ph/9307002].
  
\bibitem{Buchmuller:2012ex} 
  W.~Buchmuller, V.~Domcke and K.~Schmitz,
  JCAP {\bf 1304}, 019 (2013)
  doi:10.1088/1475-7516/2013/04/019
  [arXiv:1210.4105 [hep-ph]].

\bibitem{Buchmuller:2013zfa} 
  W.~Buchmuller, V.~Domcke and K.~Kamada,
  Phys.\ Lett.\ B {\bf 726}, 467 (2013)
  doi:10.1016/j.physletb.2013.08.042
  [arXiv:1306.3471 [hep-th]].
  
\bibitem{Einhorn:2009bh} 
  M.~B.~Einhorn and D.~R.~T.~Jones,
  JHEP {\bf 1003}, 026 (2010)
  doi:10.1007/JHEP03(2010)026
  [arXiv:0912.2718 [hep-ph]].
  
\bibitem{Kallosh:2010ug} 
  R.~Kallosh and A.~Linde,
  JCAP {\bf 1011}, 011 (2010)
  doi:10.1088/1475-7516/2010/11/011
  [arXiv:1008.3375 [hep-th]].


 \bibitem{Ferrara:2010yw} 
  S.~Ferrara, R.~Kallosh, A.~Linde, A.~Marrani and A.~Van Proeyen,
  Phys.\ Rev.\ D {\bf 82}, 045003 (2010)
  doi:10.1103/PhysRevD.82.045003
  [arXiv:1004.0712 [hep-th]].

\bibitem{Ferrara:2010in} 
  S.~Ferrara, R.~Kallosh, A.~Linde, A.~Marrani and A.~Van Proeyen,
  Phys.\ Rev.\ D {\bf 83}, 025008 (2011)
  doi:10.1103/PhysRevD.83.025008
  [arXiv:1008.2942 [hep-th]].

\bibitem{Starobinsky:1980te} 
  A.~A.~Starobinsky,
  Phys.\ Lett.\  {\bf 91B}, 99 (1980)
  doi:10.1016/0370-2693(80)90670-X;
  Sov.\ Astron.\ Lett.\  {\bf 9}, 302 (1983).

\bibitem{Binetruy:1996xj} 
  P.~Binetruy and G.~R.~Dvali,
  Phys.\ Lett.\ B {\bf 388}, 241 (1996)
  doi:10.1016/S0370-2693(96)01083-0
  [hep-ph/9606342].

\bibitem{Halyo:1996pp} 
  E.~Halyo,
  Phys.\ Lett.\ B {\bf 387}, 43 (1996)
  doi:10.1016/0370-2693(96)01001-5
  [hep-ph/9606423].

\bibitem{Kallosh:2003ux} 
  R.~Kallosh and A.~D.~Linde,
  JCAP {\bf 0310}, 008 (2003)
  doi:10.1088/1475-7516/2003/10/008
  [hep-th/0306058].

\bibitem{Kawasaki:2000yn} 
  M.~Kawasaki, M.~Yamaguchi and T.~Yanagida,
  Phys.\ Rev.\ Lett.\  {\bf 85}, 3572 (2000)
  doi:10.1103/PhysRevLett.85.3572
  [hep-ph/0004243].

\bibitem{Buchmuller:2014rfa} 
  W.~Buchmuller, V.~Domcke and K.~Schmitz,
  JCAP {\bf 1411}, no. 11, 006 (2014)
  doi:10.1088/1475-7516/2014/11/006
  [arXiv:1406.6300 [hep-ph]].
  
\bibitem{Buchmuller:2014dda} 
  W.~Buchmuller and K.~Ishiwata,
  Phys.\ Rev.\ D {\bf 91}, no. 8, 081302 (2015)
  doi:10.1103/PhysRevD.91.081302
  [arXiv:1412.3764 [hep-ph]].

\bibitem{Planck:2013jfk} 
  P.~A.~R.~Ade {\it et al.} [Planck Collaboration],
  Astron.\ Astrophys.\  {\bf 571}, A22 (2014)
  doi:10.1051/0004-6361/201321569
  [arXiv:1303.5082 [astro-ph.CO]].
  
\bibitem{Kallosh:2013yoa} 
  R.~Kallosh, A.~Linde and D.~Roest,
  JHEP {\bf 1311}, 198 (2013)
  doi:10.1007/JHEP11(2013)198
  [arXiv:1311.0472 [hep-th]].

\bibitem{Kallosh:2013xya} 
  R.~Kallosh and A.~Linde,
  JCAP {\bf 1306}, 028 (2013)
  doi:10.1088/1475-7516/2013/06/028
  [arXiv:1306.3214 [hep-th]].

\bibitem{Binetruy:2004hh} 
  P.~Binetruy, G.~Dvali, R.~Kallosh and A.~Van Proeyen,
  Class.\ Quant.\ Grav.\  {\bf 21}, 3137 (2004)
  doi:10.1088/0264-9381/21/13/005
  [hep-th/0402046].

\bibitem{Komargodski:2010rb} 
  Z.~Komargodski and N.~Seiberg,
  JHEP {\bf 1007}, 017 (2010)
  doi:10.1007/JHEP07(2010)017
  [arXiv:1002.2228 [hep-th]].

\bibitem{Dienes:2009td} 
  K.~R.~Dienes and B.~Thomas,
  Phys.\ Rev.\ D {\bf 81}, 065023 (2010)
  doi:10.1103/PhysRevD.81.065023
  [arXiv:0911.0677 [hep-th]].

\bibitem{Catino:2011mu} 
  F.~Catino, G.~Villadoro and F.~Zwirner,
  JHEP {\bf 1201}, 002 (2012)
  doi:10.1007/JHEP01(2012)002
  [arXiv:1110.2174 [hep-th]].

\bibitem{Wieck:2014xxa} 
  C.~Wieck and M.~W.~Winkler,
  Phys.\ Rev.\ D {\bf 90}, no. 10, 103507 (2014)
  doi:10.1103/PhysRevD.90.103507
  [arXiv:1408.2826 [hep-th]].

\bibitem{Domcke:2014zqa} 
  V.~Domcke, K.~Schmitz and T.~T.~Yanagida,
  Nucl.\ Phys.\ B {\bf 891}, 230 (2015)
  doi:10.1016/j.nuclphysb.2014.12.007
  [arXiv:1410.4641 [hep-th]].
  
\bibitem{Einhorn:2012ih} 
  M.~B.~Einhorn and D.~R.~T.~Jones,
  JCAP {\bf 1211}, 049 (2012)
  doi:10.1088/1475-7516/2012/11/049
  [arXiv:1207.1710 [hep-ph]].

\bibitem{Asaka:2001ez} 
  T.~Asaka, W.~Buchmuller and L.~Covi,
  Phys.\ Lett.\ B {\bf 510}, 271 (2001)
  doi:10.1016/S0370-2693(01)00623-2
  [hep-ph/0104037].

\bibitem{Whitt:1984pd} 
  B.~Whitt,
  Phys.\ Lett.\  {\bf 145B}, 176 (1984).
  doi:10.1016/0370-2693(84)90332-0

  
\end{thebibliography}
\end{document}